\documentclass[conference]{IEEEtran}
\IEEEoverridecommandlockouts
\usepackage{cite}
\usepackage{amsmath,amssymb,amsfonts}
\usepackage{algorithmic}
\usepackage{graphicx}
\usepackage{textcomp}
\usepackage{xcolor}
\usepackage{amsthm}
\usepackage{epsfig} 
\usepackage{epstopdf}
\usepackage{amsmath}
\usepackage{amsthm}
\usepackage{comment}
\usepackage{url}
\usepackage{algorithm}
\usepackage{subfigure}
\usepackage{booktabs}

\usepackage{amsfonts}
\usepackage{bm}
\usepackage{color}

\usepackage{stfloats}
\usepackage{chemarrow}
\usepackage{extarrows}
\usepackage{setspace}
\allowdisplaybreaks

\newboolean{showcomments}
\setboolean{showcomments}{true}
\newcommand{\wang}[1]{\ifthenelse{\boolean{showcomments}}
	{ \textcolor[rgb]{1,0,1}{(ZW:  #1)}}{}}
\newcommand{\fliu}[1]{\ifthenelse{\boolean{showcomments}}
	{ \textcolor{blue}{(FL:  #1)}}{}}
\newcommand{\ychen}[1]{\ifthenelse{\boolean{showcomments}}
	{ \textcolor{green}{(ZP:  #1)}}{}}
\newcommand{\slow}[1]{\ifthenelse{\boolean{showcomments}}
	{ \textcolor{blue}{(SL:  #1)}}{}}

\theoremstyle{definition}

\theoremstyle{definition}

\newtheorem{remark}{Remark}

\def\BibTeX{{\rm B\kern-.05em{\sc i\kern-.025em b}\kern-.08em
    T\kern-.1667em\lower.7ex\hbox{E}\kern-.125emX}}
\begin{document}
\setstretch{0.979}
\title{A General Initialization Scheme for Electromagnetic Transient Simulation: Towards Large-Scale Hybrid AC-DC Grids\\
\thanks{This work is supported by National Key Research and Development Program of China (2016YFB0900600) and Technology Projects of State Grid Corporation of China (52094017000W). (Corresponding author: Chen Shen)}
}

\author{\IEEEauthorblockN{Ye Liu$^{1}$, Yankan Song$^{1,2}$, Le Zhao$^{3}$, Ying Chen$^{1,2}$, Chen Shen$^{*1,2}$}
\IEEEauthorblockA{$^1$\textit{Dept. of Electrical Engineering, Tsinghua University}, Beijing, China\\
\textit{$^2$Research Center of Cloud Simulation and Intelligent Decision-Making, EIRI, Tsinghua University}, Beijing, China\\
\textit{$^3$State Grid Shanghai Electric Power Company}, Shanghai, China\\
shenchen@mail.tsinghua.edu.cn, liuye18@mails.tsinghua.edu.cn}
}

\maketitle

\begin{abstract}
With the large-scale hybrid AC-DC grids coming into being, electromagnetic transient (EMT) simulation is required to accurately describe the dynamics of systems. However, the EMT steady-state initialization for hybrid AC-DC system is difficult and time-consuming when the system scale is huge. In order to provide a stable snapshot for EMT simulation with nonlinear components and black-box components, this paper proposes a general initialization scheme for EMT simulation (EMT-GIS) which can be implemented in the electromagnetic transient program (EMTP)-type simulators. First, an integrated power flow (IPF) algorithm is introduced to provide the steady-state results. Then, an initialized snapshot calculation-and-splicing mechanism is designed for EMT-GIS. The proposed EMT-GIS is tested using a hybrid AC-DC system in China on the CloudPSS simulation platform. Test results verify the effectiveness of the proposed EMT-GIS.
\end{abstract}


\section{Introduction}
With the continuous development of power electronics technologies, traditional AC grids are gradually transformed into large-scale hybrid AC-DC grids \cite{wang2013harmonizing}, \cite{bilodeau2016making}. Due to the interactive coupling between AC and DC subsystems, the complicated dynamics of power electronics and possible cascading failures, conventional electromechanical transient simulation or electromechanical-electromagnetic hybrid simulation cannot satisfy the accurate transient analyses for hybrid AC-DC grids. Therefore, fully electromagnetic transient (EMT) simulation are highly required \cite{mahseredjian2009simulation}, \cite{watson2003power}.

The initialization process, as a prerequisite for further dynamic analyses, is one of the most important and challenging parts of EMT simulation for large-scale hybrid AC-DC systems \cite{perkins1995nonlinear}. Existing commercial EMT simulation platforms usually provide the default zero-state ramping scheme for EMT initialization \cite{faruque2005detailed}. As a simple but effective approach, the ramping scheme can rapidly initialize the small-scale EMT model. However, the initialization efficiency of the ramping scheme turns out to be low when the system scale is huge, and the complex dynamics and control logics in hybrid AC-DC systems may lead to ramping initialization failures. 

Another popular initialization scheme for EMT simulation is the steady-state initialization scheme, which initializes the EMT model directly from steady-state with the given snapshot. In \cite{noda2011practical}, a steady-state initialization method considering the harmonic power flow is proposed. In \cite{liu2017initialization}, a transition-state-calculation-based initialization approach is proposed. However, the above two methods cannot be applied to the hybrid AC-DC system because the state variables of DC's control system cannot be completely derived from the power flow results. Towards HVDC systems, an automatic EMT initialization method for LCC-HVDC is introduced in \cite{chen2018electromagnetic} and an EMT initialization scheme for MMC-HVDC is proposed in \cite{stepanov2018initialization}. Nevertheless, the EMT steady-state initialization for hybrid AC-DC system is also not considered in both \cite{chen2018electromagnetic} and \cite{stepanov2018initialization}. 

In addition, when performing bulk-grid-level simulation, there exist various black-box components (e.g. factory private model, hardware-in-the-loop device) \cite{stankovic2004transient} whose detailed models are unavailable because of the privacy protection. Obviously, the black-box components also bring difficulties to the EMT initialization and simulation of hybrid AC-DC systems, and there is no relevant research so far.

Therefore, one critical issue is how to initialize the EMT model of large-scale hybrid AC-DC system with nonlinear components and black-box components effectively. In order to address this issue, a general initialization scheme for EMT simulation (EMT-GIS) is proposed in this paper, and an integrated power flow (IPF) algorithm is proposed to provide the power flow results of the entire grid. Then, an initialized snapshot calculation-and-splicing mechanism is introduced to force the whole system run into steady-state rapidly.

The contributions of this paper are as follows.

\begin{itemize}
	\item A general scheme for EMT initialization (EMT-GIS) is proposed, which is effective for the large-scale hybrid AC-DC system cases.
	\item An IPF algorithm with the Jacobian-Free Newton-GMRES(m) (JFNG(m)) method is proposed, which provides the whole-system power flow results without the knowledge of modeling details of the black-box components.
	\item An initialized snapshot calculation-and-splicing mechanism is introduced, which improves the initialization efficiency of EMT-GIS.
\end{itemize}


The rest of this paper is organized as follows. Section II introduces the overall framework of EMT-GIS. Section III proposes the IPF algorithm with a JFNG(m)-based approach. Section IV proposes the initialized snapshot calculation-and-splicing mechanism. In section V, a hybrid AC-DC system in China is tested and the effectiveness of the proposed EMT-GIS is verified. Section VI provides the conclusion.

\section{Overall Framework of EMT-GIS}
The EMT-GIS is designed to initialize the EMT model of large-scale hybrid AC-DC systems. In a hybrid AC-DC system, subsystems with numerous power electronics (e.g. HVDC systems, renewable energy systems) are embedded in the conventional three-phase AC system (i.e. the main system). 

Considering the model availability of the components, we define two kinds of components, i.e., the \textit{black-box components} whose detailed models are hidden or private (e.g. factory private model, hardware-in-the-loop device), the \textit{white-box components} whose detailed models are available. Then, we can define the region composed of black-box components (RBC) and that composed of white-box components (RWC). The RBCs and subsystems with complex dynamics (e.g. the control dynamics of HVDC) compose the generalized region of black-box components (GRBC).
The proposed EMT-GIS deals with the main system and GRBCs by different approaches.

\begin{figure}[ht]
	\centering
	\includegraphics[width=0.47\textwidth]{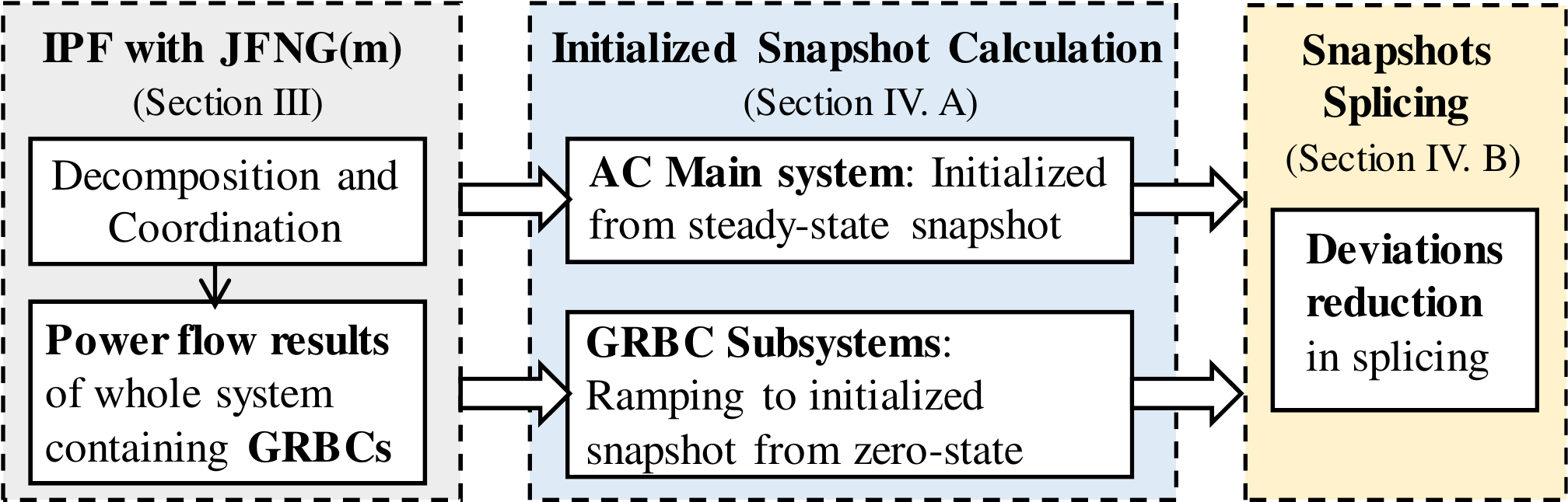}
	\caption{Overall Framework of EMT-GIS.}
	\label{scheme_gis}
\end{figure}

The framework of EMT-GIS is shown in Fig. \ref{scheme_gis}. Firstly, the proposed IPF algorithm, which considers the GRBC subsystems, provides the whole-system power flow results for EMT-GIS. Then, in the simulation process, the AC main system is initialized from the steady-state snapshot which is calculated from the IPF results, while the GRBC subsystems are initialized from zero-state by ramping approach. After all the subsystems reaching their initialized snapshots, a snapshots-splicing mechanism is introduced to splice the snapshots of each subsystem and complete the EMT initialization. Benefit from the IPF algorithm and snapshot-calculation-and-splicing mechanism, EMT-GIS has the following characteristics:

\begin{itemize}
	\item \textbf{Model-free characteristic:} the EMT-GIS is applicable to the large-scale hybrid AC-DC systems containing complex-dynamics subsystems, black-box components or even practical physical models.
	\item \textbf{Platform-free characteristic:} the EMT-GIS can be implemented in any EMT simulation platform, and in this paper, we take the cloud-computing based EMT simulation platform (CloudPSS) \cite{song2019cloudpss} as a testbed.
\end{itemize}

\section{Decomposition-and-Coordination-Based Integrated Power Flow Algorithm for EMT-GIS}
In this section, we first introduce the details of decomposition and coordination in the hybrid AC-DC system with black-box components. Then, a JFNG(m)-based IPF algorithm is proposed to obtain the power flow results of the whole system.
 
\subsection{Decomposition and Coordination}

When decomposing the whole hybrid AC-DC system, the RWC system can be considered as the main system with effective power flow solver (e.g. AC-DC power flow solver for HVDC systems), or as the GRBC system if the detailed model is too complex to solve. Thus, in this section, we consider the decomposition and coordination among the main system and GRBC systems.

\begin{figure}[h]
	\centering
	\setlength{\abovecaptionskip}{-0.05cm}   
	\includegraphics[width=0.43\textwidth]{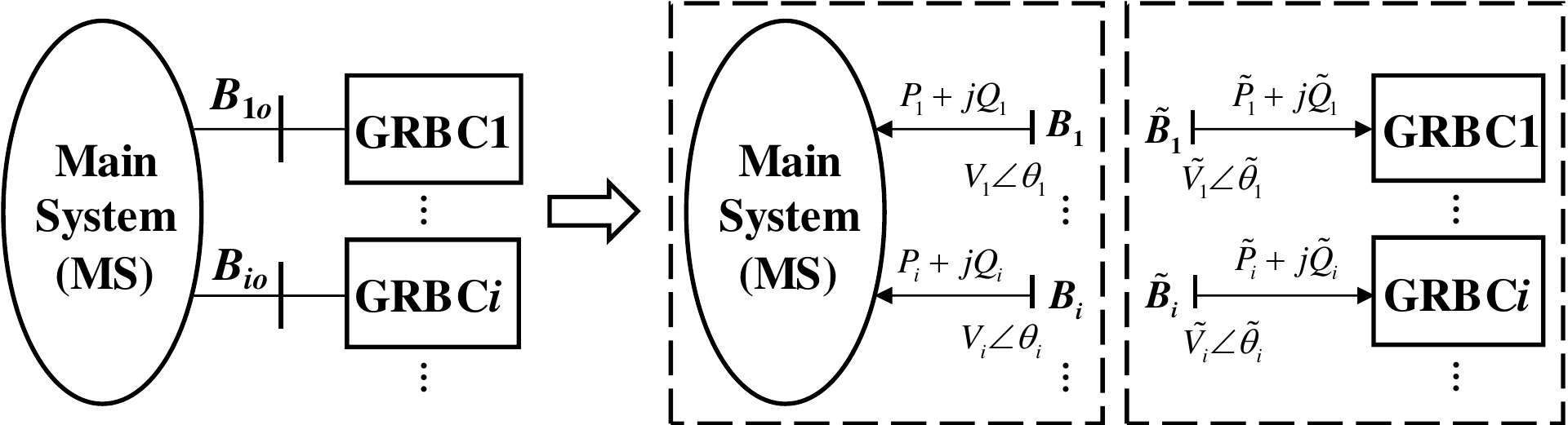}
	\caption{Topology and decomposition of the whole system.}
	\label{de_co_fig}
\end{figure}

Generally, a system decomposition is shown in Fig. \ref{de_co_fig}, where the main system is connected to GRBC systems through boundary buses. The set of GRBCs is denoted by $\mathcal{N}_B=\{1,2,\dots,n\}$, which can also represent the boundary buses.

The decomposition is achieved by boundary buses tearing. For $i \in \mathcal{N}_B$, the boundary bus $B_{io}$ is torn into $\tilde{B}_i$ for GRBC$i$ and $B_i$ for main system respectively, and the corresponding injected power and voltage of each system can be represented as $(\tilde{P}_i +j \tilde{Q}_i, \tilde{V}_i \angle \tilde{\theta}_i)$ and $(P_i +j Q_i, V_i \angle \theta_i)$, as shown in Fig. \ref{de_co_fig}.
Supposing the boundary voltages $\tilde{V}_i \angle \tilde{\theta}_i$ and $V_i \angle \theta_i$ are provided, the power flow of the main system can be solved by setting the boundary buses as slack buses, and the injected power $\tilde{P}_i +j \tilde{Q}_i$ of GRBC$i$ system can be obtained by digital simulation tools or practical physical simulation. 

Obviously, to derive the power flow results of the whole system, it is necessary for the systems' variables to satisfy the boundary convergence condition, which is written in the vector form as follows:
\begin{subequations}
	\label{boundary_conv_con}
	\setlength{\abovedisplayskip}{1pt}
	\begin{align}
	\bm{V}=\tilde{\bm{V}}&,\bm{\theta}=\tilde{\bm{\theta}}
	\label{boundary_conv_con_1}
	\\
	\bm{P}+\tilde{\bm{P}}=\bm{0}&,\bm{Q}+\tilde{\bm{Q}}=\bm{0}
	\end{align}
\end{subequations}  
where $\bm{V}=\text{col}(V_i)$, $\tilde{\bm{V}}=\text{col}(\tilde{V}_i)$, $\bm{\theta}=\text{col}(\theta_i)$, $\tilde{\bm{\theta}}=\text{col}(\tilde{\theta}_i)$, $\bm{P}=\text{col}(P_i)$, $\tilde{\bm{P}}=\text{col}(\tilde{P}_i)$, $\bm{Q}=\text{col}(Q_i)$ and $\tilde{\bm{Q}}=\text{col}(\tilde{Q}_i) \in \mathbb{R}^n$ for $i \in \mathcal{N}_B$. 

\begin{figure}[h]
	\centering
	\setlength{\abovecaptionskip}{-0.1cm}   
	\setlength{\belowcaptionskip}{-0.1cm}   
	\includegraphics[width=0.45\textwidth]{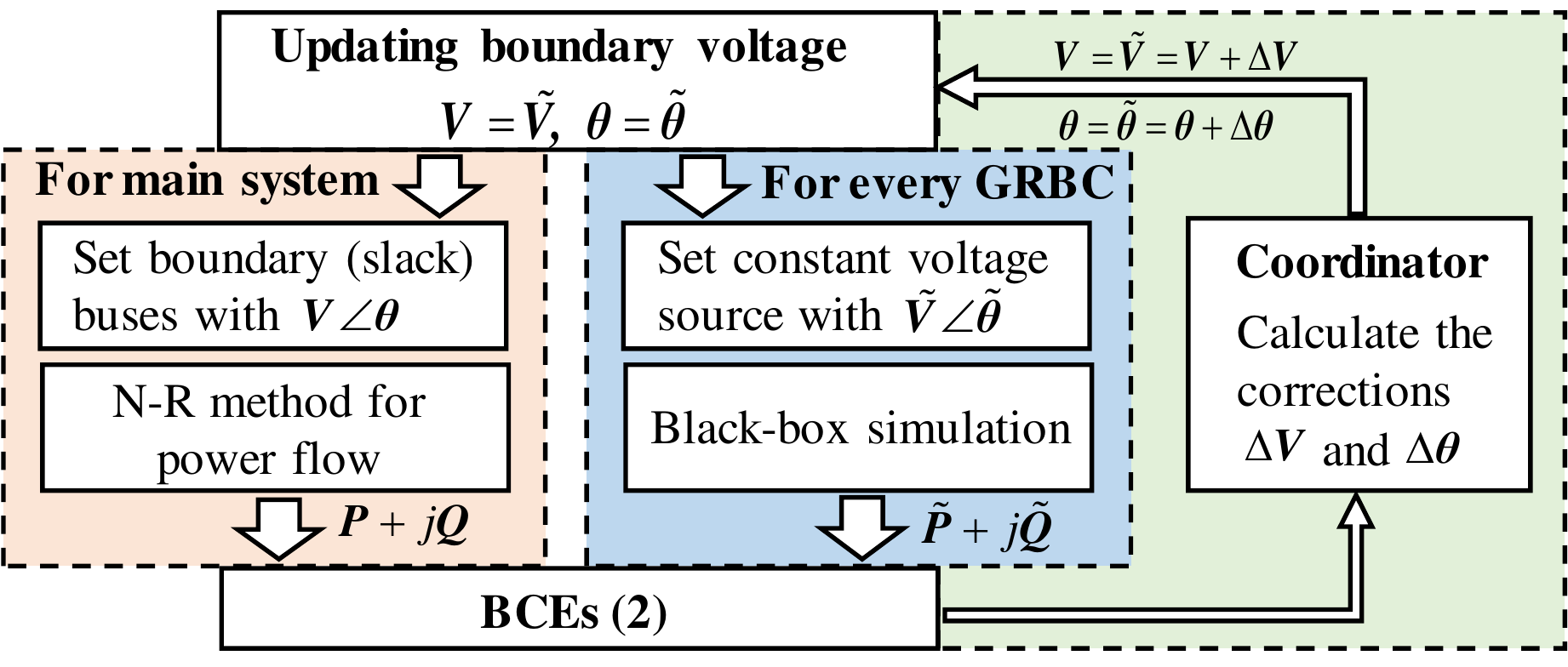}
	\caption{Procedures of the IPF.}
	\label{dncpf}
\end{figure}

If the convergence condition \eqref{boundary_conv_con} is not satisfied completely, with the given equal boundary voltages which satisfy \eqref{boundary_conv_con_1}, the boundary power deviation vectors are $\Delta\bm{P}=\bm{P}+\tilde{\bm{P}}$ and $\Delta\bm{Q}=\bm{Q}+\tilde{\bm{Q}}$. Then, the boundary coordination equations (BCEs) can be expressed as \eqref{co_equ}.
\begin{align}
\label{co_equ}
\bm{\Phi}(\bm{V}^\mathrm{T},\bm{\theta}^\mathrm{T})=
\begin{bmatrix}
\Delta\bm{P}\\\Delta\bm{Q}
\end{bmatrix}
=
\begin{bmatrix}
\bm{P}+\tilde{\bm{P}}\\\bm{Q}+\tilde{\bm{Q}}
\end{bmatrix}
=\bm{0}
\end{align}
where $\bm{\Phi}(\bm{V}^\mathrm{T},\bm{\theta}^\mathrm{T}) \in \mathbb{R}^{2n}$. Thus, the solution of the whole system's power flow is equivalent to the solution of BCEs \eqref{co_equ}, and the procedures of integrated power flow (IPF) are as shown in Fig. \ref{dncpf}.

\begin{remark}
	The power flow results from IPF benefit the EMT-GIS. The BCEs are implicit functions of $\bm{V}$ and $\bm{\theta}$. However, the coordinator for obtaining the corrections in Fig. \ref{dncpf} is difficult to design because of the unavailability of detailed internal models of the GRBC systems, thus the Jacobian matrix of the whole system cannot be derived during the Newton-Raphson solving process. To solve the problem aforementioned, the Jacobian-Free Newton-GMRES(m) (JFNG(m)) algorithm is utilized to solve the BCEs in the following subsection.
\end{remark}

\subsection{JFNG(m) Method for Boundary Coordination}

JFNG(m) is a nonlinear equations solver without explicitly forming the Jacobian matrix. An adaptive preconditioner in \cite{chen2006jacobian} is constructed for JFNG(m) method to enhance the convergence. In this subsection, the JFNG(m) method with the adaptive preconditioner is utilized to solve the BCEs in the IPF process. The details of the IPF-JFNG(m) algorithm are shown in \textbf{Algorithm 1}.

\begin{algorithm}[ht]
	\caption{IPF-JFNG(m) Algorithm} 
	\label{DnCPF_Alg}
	\small
	\hangafter 1
	\hangindent 1em
	\textbf{Input:} Initial boundary voltage $\bm{V}^0=\tilde{\bm{V}}^0$, $\bm{\theta}^0=\tilde{\bm{\theta}}^0$ and initial vector $\bm{x}^0=((\bm{V}^0)^{\mathrm{T}},(\bm{\theta}^0)^{\mathrm{T}}) \in \mathbb{R}^{2n}$, very small error tolerances $\varepsilon_1, \varepsilon_2>0$, initial non-singular preconditioning matrix $\bm{M}^0$, and iteration index $k=0$.
	
	\hangafter 1
	\hangindent 1em
	\textbf{Output:} power flow results of the whole system.
	
	\hangafter 1
	\hangindent 1em
	\textbf{Step 1:} Derive $\bm{P}^k$, $\bm{Q}^k$ and $\tilde{\bm{P}}^k$, $\tilde{\bm{Q}}^k$ as shown in Fig. \ref{dncpf}.
	
	\hangafter 1
	\hangindent 1em
	\textbf{Step 2:} Form the boundary coordination equations \eqref{co_equ} and calculate the Euclidean norm $\|\bm{\Phi}(\bm{x}^k)\|_2$.
	
	\hangafter 1
	\hangindent 1em
	\textbf{If} $\|\bm{\Phi}(\bm{x}^k)\|_2 \le \varepsilon_1$, the power flow of the whole system converges, \textbf{exit}.
	
	\hangafter 1
	\hangindent 1em
	\textbf{else} continue \textbf{Step 3} - \textbf{Step 5}.
	
	\hangafter 1
	\hangindent 1em
	\textbf{Step 3:} Utilize GMRES(m) algorithm \cite{saad1986gmres} to iteratively solve the correction equation: 
	\begin{align}
		\bm{\Phi}^{\prime}(\bm{x}^k) \Delta \bm{x}^k = -\bm{\Phi}(\bm{x}^k)
	\end{align}
	
	\hangafter 0
	\hangindent 1em
	\textbf{S 3.1:} Set $\bm{r}_0 = -\bm{\Phi}(\bm{x}^k)$, $l=1$, $\rho=\beta=\|\bm{r}_0\|_2$, $\bm{v}^1=\bm{r}_0$, $\varepsilon_G=\varepsilon_2 \|\bm{r}_0\|_2>0$.
	
	\hangafter 0
	\hangindent 1em
	\textbf{S 3.2:} Let $\bm{z}^l=\bm{M}^k \bm{v}^l$, calculate the variables $\Delta \bm{\Phi}^l = \bm{\Phi}(\bm{x}^k+\omega \bm{z}^l)-\bm{\Phi}(\bm{x}^k)$, $\Delta \bm{x}^l=\omega \bm{z}^l$, and $\bm{v}^{l+1}=\frac{\Delta\bm{\Phi}^l}{\omega}$.
	
	\hangafter 0
	\hangindent 1em
	\textbf{S 3.3:} Correct the preconditioning matrix by \eqref{algo_2}.
	\begin{align}
	\label{algo_2}
	\bm{M}^{k}=\bm{M}^{k}+\frac{\left(\Delta \bm{x}^{l}-\bm{M}^{k} \Delta \bm{\Phi}^{l}\right)\left(\Delta \bm{x}^{l}\right)^{\mathrm{T}} \bm{M}^{k}}{\left(\Delta \bm{x}^{l}\right)^{\mathrm{T}} \bm{M}^{k} \Delta \bm{\Phi}^{l}}
	\end{align}
	
	\hangafter 0
	\hangindent 1em
	\textbf{S 3.4:} Orthogonalize $\bm{V}^{l+1}=(\bm{v}^1,\bm{v}^2,\dots,\bm{v}^{l+1})$ and then derive the Hessian matrix $\overline{\bm{H}}^l$. Obtain $\rho$ and $\bm{y}^l$ by solving the optimization $\rho=\min\limits_{\bm{y}^l \in \mathbb{R}^l}\left( \left\| \beta\bm{e}_1 - \overline{\bm{H}}^l \bm{y}^l\right\| \right)$. 
	
	\hangafter 0
	\hangindent 1em
	\textbf{If} $\rho \le \varepsilon_G$, we obtain $\Delta \bm{x}^k=\bm{M}^k \bm{V}^l \bm{y}^l$.
	
	\hangafter 0
	\hangindent 1em
	\textbf{else} set $l=l+1$ where $l<m$, reture to \textbf{S 3.2}.
	
	\hangafter 1
	\hangindent 1em
	\textbf{Step 4:} Calculate $\bm{x}^{k+1}=\bm{x}^{k}+\Delta \bm{x}^k$, $\Delta \bm{\Phi}^k=\bm{\Phi}(\bm{x}^{k+1})-\bm{\Phi}(\bm{x}^{k})$ 
	
	\hangafter 1
	\hangindent 1em
	\textbf{Step 5:} Correct the preconditioning matrix by \eqref{algo_3}, set $k=k+1$ and return to \textbf{Step 1}.
	\begin{align}
	\label{algo_3}
	\bm{M}^{k+1}=\bm{M}^{k}+\frac{\left(\Delta \bm{x}^{k}-\bm{M}^{k} \Delta \bm{\Phi}^{k}\right)\left(\Delta \bm{x}^{k}\right)^{\mathrm{T}} \bm{M}^{k}}{\left(\Delta \bm{x}^{k}\right)^{\mathrm{T}} \bm{M}^{k} \Delta \bm{\Phi}^{k}}
	\end{align}
	
\end{algorithm}  

\begin{remark}
	The power flow results of the IPF-JFNG(m) algorithm enable the EMT initialization of the large-scale hybrid AC-DC system containing black-box components. The convergence of the IPF-JFNG(m) algorithm is equivalently discussed in \cite{chen2006jacobian} and \cite{saad1986gmres}.
\end{remark}

\section{Initialized Snapshot Calculation-and-Splicing Mechanism}
In this section, an initialized snapshot calculation-and-splicing mechanism is developed, which is also a significant part of EMT-GIS. Firstly, each subsystem calculates its initialized snapshot according to the IPF results respectively. Then, a snapshots-splicing approach is introduced. 

\subsection{Initialized Snapshot Calculation}
We introduce the initialized snapshot calculation in the following two parts.

\textit{1) Three-Phase AC Main System}. The main system consists of three-phase AC components (e.g. synchronous generators and three-phase transformers). Considering the scenarios with three-phase symmetrical operation, the electrical variables can be expressed in the phasor form. Thus, given the steady-state power flow data, we can initialize the internal state variables of the components by the phasor diagram calculation approach, which enables the EMT initialization directly from steady-state. 

Supposing the power flow results of component $g$ are $V_g^{pf}$, $\theta_g^{pf}$, $P_g^{pf}$ and $Q_g^{pf}$, which can be further expressed as port voltage $\dot{V}_g^{pf}=V_g^{pf} \angle \theta_g^{pf}$ and injection power $\dot{S}_g^{pf}=P_g^{pf} +j Q_g^{pf}$. Then, the current phasor is $\dot{I}_g^{pf}=(\dot{S}_g^{pf}/\dot{V}_g^{pf} )^*$, and the phasors of other state variables (e.g. the transient potentials of synchronous generators) can be derived from the phasor diagram calculation \cite{kundur1994power}. Further, we calculate the instantaneous values of state variables at time $t$ to compose the initialized snapshot at time $t$. Take current for example, the instantaneous current $i_g(t)$ of component $g$ can be differenced and represented as \eqref{ins_cur}.
\begin{align}
i_g(t)=G v_g(t)+ H v_g(t-\Delta t)+J i_g(t-\Delta t)
\label{ins_cur}
\end{align}
where $v_g(t)$ is the instantaneous voltage, $G$, $H$, $J$ are the equivalent parameters determined by $\Delta t$ and component parameters, $v_g(t-\Delta t)$ and $i_g(t-\Delta t)$ are the historical voltage and current before $\Delta t$, which can be represented by the phasors as \eqref{phasor_ins}
\begin{subequations}
	\label{phasor_ins}
	\setlength{\abovedisplayskip}{1.5pt}
	\setlength{\belowdisplayskip}{1.5pt}
	\begin{align}
	i_g(t-\Delta t)&=\text{Re} \left(\dot{I}_g^{pf} e^{j\omega (t-\Delta t)}\right)\\
	v_g(t-\Delta t)&=\text{Re} \left(\dot{V}_g^{pf} e^{j\omega (t-\Delta t)}\right)
	\end{align}
\end{subequations}

Then, after the calculation of all instantaneous values of the state variables, we can derive the initialized snapshot of the AC main system, where the main system can be initialized directly from.

\textit{2) GRBC Subsystems}. As the definition in Section II, GRBC subsystems consist of RBCs and subsystems with complex dynamics (e.g. HVDC systems), in which the initialized internal variables cannot be calculated directly by the power flow results. For RBCs, the reasons are the unavailability of the detailed models. And for subsystems with complex dynamics, the amount of unknown variables is beyond that of the solving equations, thus it is also difficult to derive the initialized snapshot. To solve the aforementioned problem, a ramping approach with ideal sources is utilized to lead the EMT models of GRBCs to initialized snapshots.

\begin{figure}[ht]  
	\centering
	\vspace{-0.05cm}  
	\setlength{\abovecaptionskip}{-0.05cm}   
	\setlength{\belowcaptionskip}{-0.1cm}   
	\subfigure[Steady-state operation]{\includegraphics[width=1.5in]{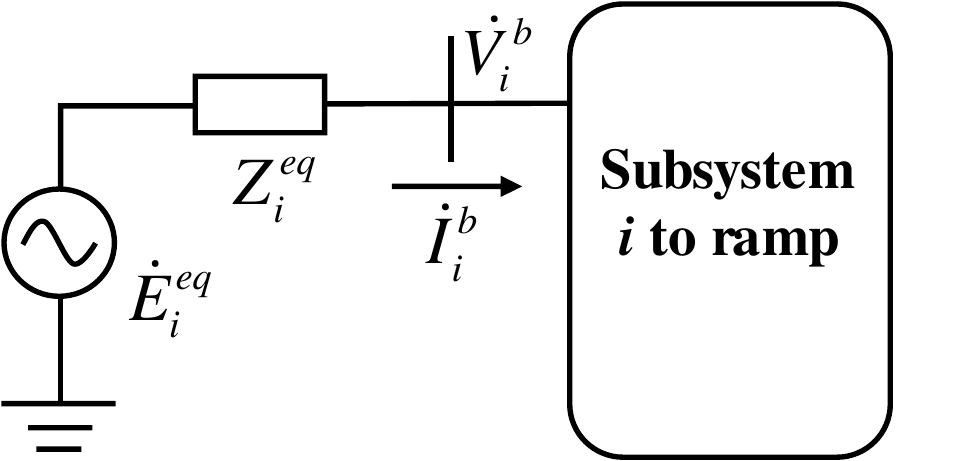}}  \subfigure[Fault analysis]{\includegraphics[width=1.5in]{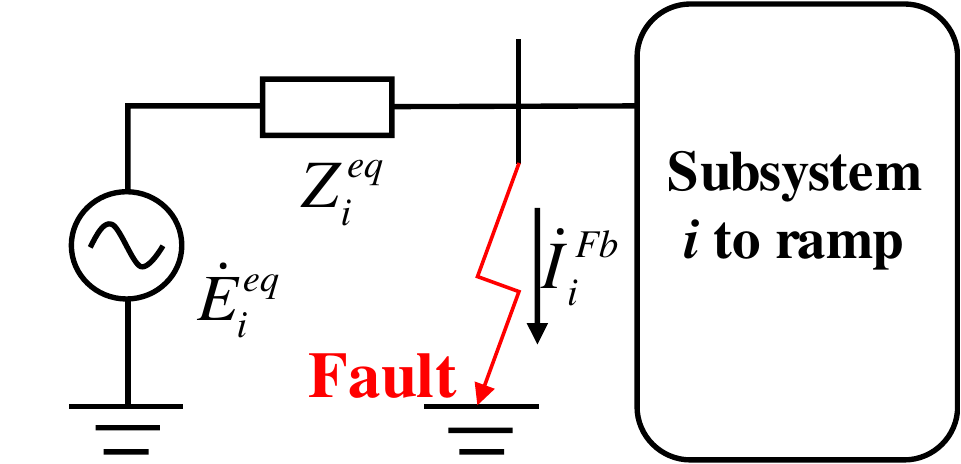}}  
	\caption{Thevenin equivalent calculation.}  
	\label{thevenin_eq}
\end{figure}

For GRBC subsystem $i$, to simulate the characteristic when this subsystem is connected to the main system, the Thevenin equivalent of the remaining system is required to ramp up the subsystem. In the simulation environment, the Thevenin equivalent can be derived from power flow data and a fault analysis at the boundary \cite{huang2019transient}, as shown in Fig. \ref{thevenin_eq}.


We derive \eqref{theve_equ_1} from Fig. \ref{thevenin_eq}(a) and \eqref{theve_equ_2} from Fig. \ref{thevenin_eq}(b).
\begin{subequations}
	\setlength{\belowdisplayskip}{0.5pt}
	\begin{align}
	&\dot{E}_i^{eq}=\dot{I}_i^b Z_i^{eq} + \dot{V}_i^b
	\label{theve_equ_1}\\
	&\dot{E}_i^{eq}=\dot{I}_i^{Fb} Z_i^{eq}
	\label{theve_equ_2}
	\end{align}
\end{subequations}
where $\dot{E}_i^{eq}$ is the Thevenin source, $Z_i^{eq}$ is the Thevenin impedance, $\dot{V}_i^b$ and $\dot{I}_i^b$ are the boundary voltage and current of steady-state operation, and $\dot{I}_i^{Fb}$ is the fault current to ground. Combining \eqref{theve_equ_1} and \eqref{theve_equ_2}, we have:
\begin{align}
\setlength{\abovedisplayskip}{0pt}
\setlength{\belowdisplayskip}{0pt}
Z_i^{eq}=\dot{V}_i^b/(\dot{I}_i^b - \dot{I}_i^{Fb})
\end{align}

Then, during the ramping process with Thevenin equivalent sources, GRBC subsystems reach their initialized snapshots.

\subsection{Snapshots Splicing}
After every subsystem running into the initialized snapshot, the snapshots splicing mechanism interconnects all subsystems in the simulation environment, splices the snapshots of each subsystem, and then completes the EMT initialization.

There are various approaches to technically realize the snapshots splicing. However, since each subsystem reaches the calculated initialized snapshot independently and non-simultaneously, there may exist deviations among the instantaneous values of the subsystems' AC port variables caused by the phase differences of the variables, and the worst scenario is that two variables have opposite phases at splicing time. These splicing deviations will further result in splicing oscillations, and in some severe scenarios, the splicing oscillations will lead to initialization failures. Thus, one critical problem about the splicing mechanism design is how to reduce or even eliminate the splicing deviations.

Towards the above problem, a splicing time adjustment approach is introduced to reduce the splicing deviations. Generally, supposing that the variables $y_i$ of subsystem $i$ with phase $\varphi_i$ and $y_j$ of subsystem $j$ with phase $\varphi_j$ need to be spliced. These two subsystems reach their initialized snapshots at time $t_i$ and $t_j$ respectively, and we suppose that $t_i < t_j$. Obviously, the splicing deviation of instantaneous values $|y_i(t_i)-y_j(t_j)|$ is supposed to be zero if there is:
\begin{align}
\varphi_i(t_i)=\varphi_j(t_j)
\label{splic}
\end{align} 

In this splicing time adjustment approach, we adjust the slower time $t_j$, and we can select a splicing time $t_j^{adj}>t_j$ for subsystem $j$ to reduce the deviation. By \eqref{splic}, the adjusted splicing time $t_{j}^{adj}$ can be derived from $\min(t_{j}^{adj}-t_{j})$ subject to $\varphi_i(t_i)=\varphi_j(t_j)$, which is equivalent to \eqref{sub}.
\begin{align}
t_j^{adj}-t_i=2kT, \ k \in \mathbb{Z}
\label{sub}
\end{align}
where $T$ is the period of AC system. Then, with the adjusted splicing times for each subsystems, the whole system will complete initialization process efficiently.

\section{Case Study}
In this section, the accuracy and effectiveness of the proposed EMT-GIS is illustrated by a case study on a hybrid AC-DC system in China.
\subsection{System Description}
\begin{figure}[ht]
	\centering
	\includegraphics[width=0.48\textwidth]{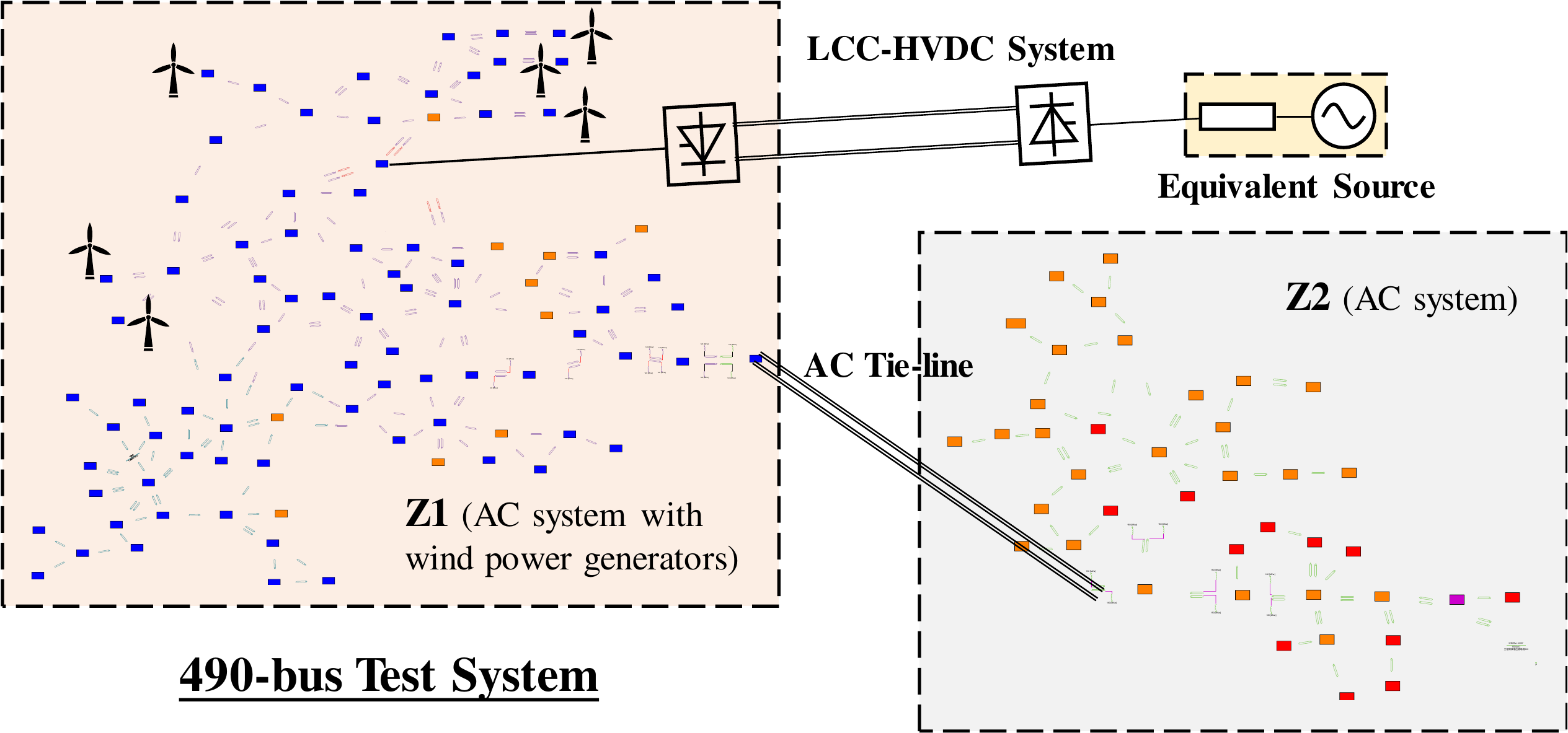}
	\caption{Topology of hybrid AC-DC test system.}
	\label{topo_case}
\end{figure}

The topology of the hybrid AC-DC test system is shown in Fig. \ref{topo_case}, which contains 490 buses in total. The full EMT model of the test system is built on the CloudPSS platform \cite{song2019cloudpss}. A $\pm$400kV bipolar 12-pulse LCC-HVDC system, which adopts the CIGRE HVDC model (white-box model), is connected to the receiving-end system Z1. Moreover, the wind turbines in Z1 adopt the GW models provided by the Goldwind company, which are black-box components because of the privacy protection of the control system. Select the DC buses and wind turbine buses as boundary buses.

We implement the proposed EMT-GIS in CloudPSS platform, and then test the initialization performance with the EMT-GIS and the default zero-state initialization scheme \cite{liu2018modeling} respectively (default steady-state initialization scheme is not tested due to its inapplicability to DC systems). Simulation results are provided in the following subsection.

\subsection{Simulation Results}

The power flow results of the test system are solved by IPF-JFNG(m) algorithm (not shown here due to space limitations), which enable the EMT-GIS.

\begin{figure}[ht]
	\centering
	\vspace{-0.1cm}  %
	\setlength{\abovecaptionskip}{0.cm}
	\setlength{\belowcaptionskip}{-0.cm}
	\includegraphics[width=0.48\textwidth]{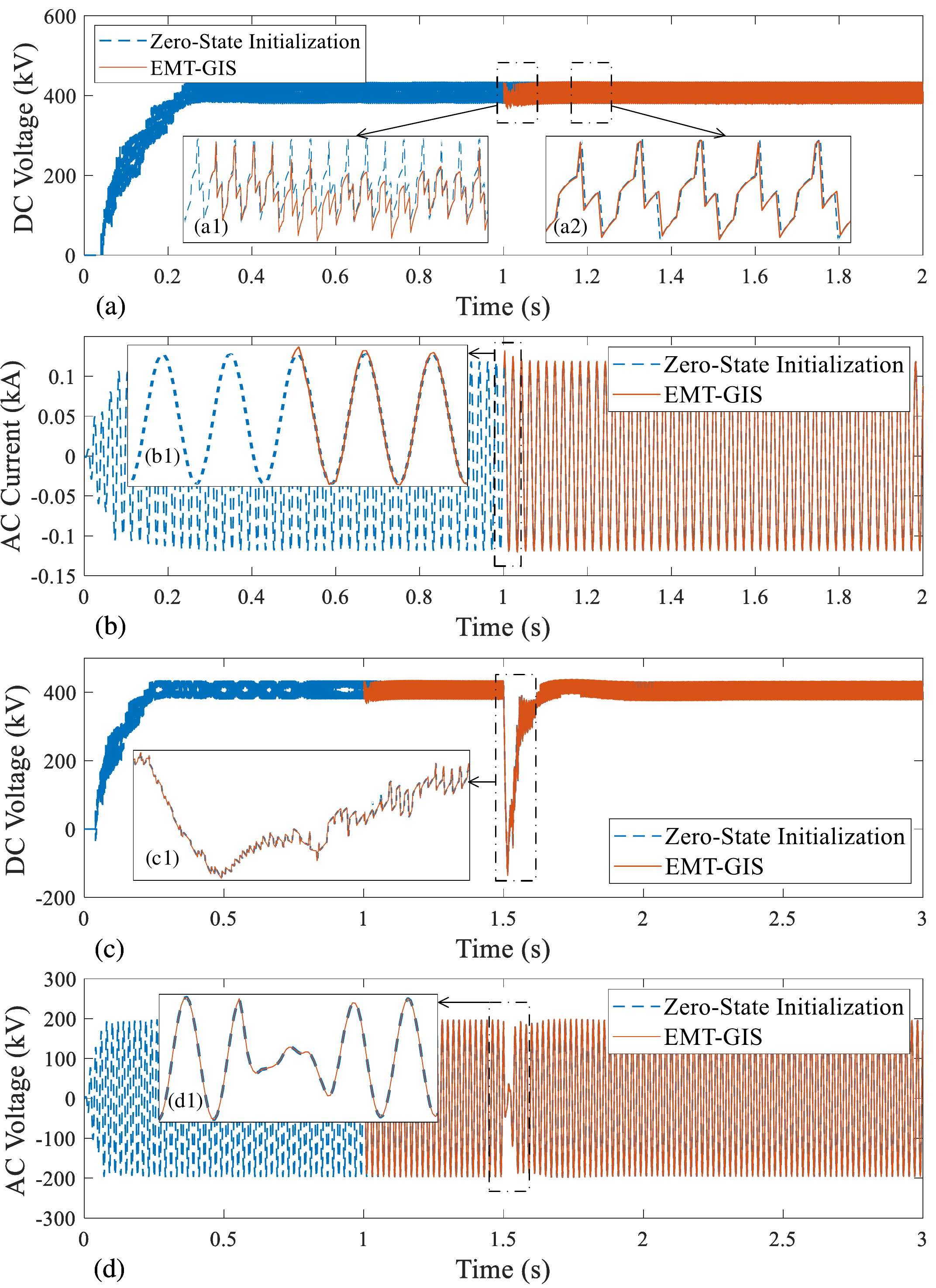}
	\caption{Simulation results. (a) DC Voltage for initialization test. (b) AC Current for initialization test. (c) DC Voltage for fault test. (d) AC Voltage for fault test.}
	\label{result_1}
\end{figure}

The accuracy test results are shown in Fig. \ref{result_1}, where we select DC voltage and AC current for initialization test and DC and AC voltages for fault test. The whole system is initialized directly from steady-state by EMT-GIS with the setting time t = 1s. As shown in Fig. \ref{result_1}(a1) and Fig. \ref{result_1}(b1), there exist tiny splicing deviations for a short time at about 1s, which are reduced by the splicing mechanism. After about 1.2s, the deviations disappear and the steady-states with two schemes tend to be identical, as shown in Fig. \ref{result_1}(a2). Fig. \ref{result_1}(c) and Fig. \ref{result_1}(d) verifies the accuracy of the transient analyses after EMT-GIS, where the fault dynamics after EMT-GIS are very close to that after zero-state ramping. Further, the average relative deviations between the above two schemes within $\Delta t=0.1s$ are shown in Table. I, which turn out to be tiny and also verify the accuracy of EMT-GIS.

\begin{table}[htbp] 
	\footnotesize
	\centering
	\vspace{-0.4cm}  %
	\setlength{\abovecaptionskip}{0.cm}
	\setlength{\belowcaptionskip}{-0.cm}
	\caption{Average Relative Deviations between EMT-GIS and Zero-State Initialization}
	\label{tab1} 
	\begin{tabular}{cccc} 
		\toprule 
		Test Type & Variable & Time Period & Deviation\\ 
		\midrule 
		Initialization Test & DC Voltage & 1.7 - 1.8s & $2.7\times10^{-3}$ \\
		Initialization Test & AC Current & 1.7 - 1.8s & $1.2\times10^{-4}$\\
		Fault Test & DC Voltage & 1.6 - 1.7s & $1.2\times10^{-2}$\\
		Fault Test & AC Voltage & 1.6 - 1.7s & $9.5\times10^{-4}$\\  
		\bottomrule 
	\end{tabular} 
\end{table}

Then, the efficiency of EMT-GIS is tested on the CloudPSS platform. With the simulation step size $\Delta T_s=20\mu s$, the initialization time of the test system with EMT-GIS is about 3.0s, while that with zero-state ramping initialization is about 44.7s. Thus, the EMT-GIS significantly improves the initialization efficiency. 

%

\section{Conclusion}
In this paper, a general initialization scheme for EMT simulation (EMT-GIS) is proposed to effectively initialize the EMT model of large-scale hybrid AC-DC grids containing black-box components. An IPF-JFNG(m) algorithm is introduced to provide the power flow results for EMT-GIS, which deals with the power flow problem of black-box components and ensures the accuracy of initialization. Then, an initialized snapshot calculation-and-splicing mechanism is proposed to ensure the initialization efficiency. the case study on a hybrid AC-DC system in China is performed on the CloudPSS platform, and the simulation results validate the accuracy and effectiveness of the proposed EMT-GIS.


\bibliographystyle{IEEEtran}
\bibliography{mybib}

\begin{thebibliography}{10}
\providecommand{\url}[1]{#1}
\csname url@samestyle\endcsname
\providecommand{\newblock}{\relax}
\providecommand{\bibinfo}[2]{#2}
\providecommand{\BIBentrySTDinterwordspacing}{\spaceskip=0pt\relax}
\providecommand{\BIBentryALTinterwordstretchfactor}{4}
\providecommand{\BIBentryALTinterwordspacing}{\spaceskip=\fontdimen2\font plus
\BIBentryALTinterwordstretchfactor\fontdimen3\font minus
  \fontdimen4\font\relax}
\providecommand{\BIBforeignlanguage}[2]{{%
\expandafter\ifx\csname l@#1\endcsname\relax
\typeout{** WARNING: IEEEtran.bst: No hyphenation pattern has been}%
\typeout{** loaded for the language `#1'. Using the pattern for}%
\typeout{** the default language instead.}%
\else
\language=\csname l@#1\endcsname
\fi
#2}}
\providecommand{\BIBdecl}{\relax}
\BIBdecl

\bibitem{wang2013harmonizing}
P.~Wang, L.~Goel, X.~Liu, and F.~H. Choo, ``Harmonizing {AC} and {DC}: A hybrid
  {AC/DC} future grid solution,'' \emph{IEEE Power Energy Mag.}, vol.~11,
  no.~3, pp. 76--83, 2013.

\bibitem{bilodeau2016making}
H.~Bilodeau, S.~Babaei, B.~Bisewski, J.~Burroughs, C.~Drover, J.~Fenn,
  B.~Fardanesh, B.~Tozer, B.~Shperling, and P.~Zanchette, ``Making old new
  again: {HVDC} and {FACTS} in the northeastern {U}nited {S}tates and
  {C}anada,'' \emph{IEEE Power Energy Mag.}, vol.~14, no.~2, pp. 42--56, 2016.

\bibitem{mahseredjian2009simulation}
J.~Mahseredjian, V.~Dinavahi, and J.~A. Martinez, ``Simulation tools for
  electromagnetic transients in power systems: Overview and challenges,''
  \emph{IEEE Trans. Power Delivery}, vol.~24, no.~3, pp. 1657--1669, 2009.

\bibitem{watson2003power}
N.~Watson, J.~Arrillaga, and J.~Arrillaga, \emph{Power systems electromagnetic
  transients simulation}.\hskip 1em plus 0.5em minus 0.4em\relax IET, 2003,
  vol.~39.

\bibitem{perkins1995nonlinear}
B.~Perkins, J.~Marti, and H.~Dommel, ``Nonlinear elements in the {EMTP}:
  steady-state initialization,'' \emph{IEEE Trans. Power Syst.}, vol.~10,
  no.~2, pp. 593--601, 1995.

\bibitem{faruque2005detailed}
M.~Faruque, Y.~Zhang, and V.~Dinavahi, ``Detailed modeling of {CIGRE} {HVDC}
  benchmark system using {PSCAD/EMTDC} and {PSB/SIMULINK},'' \emph{IEEE trans.
  Power Delivery}, vol.~21, no.~1, pp. 378--387, 2005.

\bibitem{noda2011practical}
T.~Noda and K.~Takenaka, ``A practical steady-state initialization method for
  electromagnetic transient simulations,'' in \emph{Proc. of IPST}, 2011.

\bibitem{liu2017initialization}
J.~Liu, X.~Hao, W.~Fang, Z.~Wei, T.~Wei, H.~Xu, S.~Niu, and L.~Cheng,
  ``Initialization of full electromagnetic transient simulation via a novel
  transition state calculation,'' in \emph{2017 IEEE International Conference
  on Energy Internet (ICEI)}.\hskip 1em plus 0.5em minus 0.4em\relax IEEE,
  2017, pp. 7--12.

\bibitem{chen2018electromagnetic}
X.~Chen, X.~Zhang, F.~Tian, W.~Qiu, Y.~Jin, and X.~Zhou, ``Electromagnetic
  model automatic initialization method for {HVDC} transmission system,'' in
  \emph{2018 International Conference on Power System Technology
  (POWERCON)}.\hskip 1em plus 0.5em minus 0.4em\relax IEEE, 2018, pp. 370--375.

\bibitem{stepanov2018initialization}
A.~Stepanov, H.~Saad, U.~Karaagac, and J.~Mahseredjian, ``Initialization of
  modular multilevel converter models for the simulation of electromagnetic
  transients,'' \emph{IEEE Trans. Power Delivery}, vol.~34, no.~1, pp.
  290--300, 2018.

\bibitem{stankovic2004transient}
A.~M. Stankovic and A.~T. Saric, ``Transient power system analysis with
  measurement-based gray box and hybrid dynamic equivalents,'' \emph{IEEE
  Trans. Power Syst.}, vol.~19, no.~1, pp. 455--462, 2004.

\bibitem{song2019cloudpss}
Y.~Song, Y.~Chen, Z.~Yu, S.~Huang, and C.~Shen, ``Cloud{PSS}: A
  high-performance power system simulator based on cloud computing,''
  \emph{arXiv preprint arXiv:1903.01081}, 2019.

\bibitem{chen2006jacobian}
Y.~Chen and C.~Shen, ``A {J}acobian-free {N}ewton-{GMRES}(m) method with
  adaptive preconditioner and its application for power flow calculations,''
  \emph{IEEE Trans. Power Syst.}, vol.~21, no.~3, pp. 1096--1103, 2006.

\bibitem{saad1986gmres}
Y.~Saad and M.~H. Schultz, ``G{MRES}: A generalized minimal residual algorithm
  for solving nonsymmetric linear systems,'' \emph{SIAM J. Sci. Comput.},
  vol.~7, no.~3, pp. 856--869, 1986.

\bibitem{kundur1994power}
P.~Kundur, N.~J. Balu, and M.~G. Lauby, \emph{Power system stability and
  control}.\hskip 1em plus 0.5em minus 0.4em\relax McGraw-hill New York, 1994,
  vol.~7.

\bibitem{huang2019transient}
S.~Huang, W.~Wei, Y.~Liu, Y.~Song, Y.~Chen, L.~Tang, and B.~Zhou, ``Transient
  model integrated power flow and its application in electromagnetic transient
  simulation initialization,'' in \emph{2019 IEEE Innovative Smart Grid
  Technologies-Asia (ISGT Asia)}.\hskip 1em plus 0.5em minus 0.4em\relax IEEE,
  2019, pp. 587--592.

\bibitem{liu2018modeling}
Y.~Liu, Y.~Song, Z.~Yu, C.~Shen, and Y.~Chen, ``Modeling and simulation of
  hybrid {AC-DC} system on a cloud computing based simulation
  platform-{C}loud{PSS},'' in \emph{2018 2nd IEEE Conference on Energy Internet
  and Energy System Integration (EI2)}.\hskip 1em plus 0.5em minus 0.4em\relax
  IEEE, 2018, pp. 1--6.

\end{thebibliography}


\end{document}